\documentclass[a4,twocolumn]{revtex4-1}
\usepackage{amsmath}
\usepackage{amssymb}
\usepackage{mathtools}
\usepackage{mathrsfs}
\usepackage{natbib}

\begin{document}
\bibliographystyle{numeric}
\title{Concurrence of the Blandford-Payne Process and the Bardeen-Petterson Effect: Theoretical Prediction and its Observational Evidences}
\author{Hongsu Kim\footnote{hongsu@kasi.re.kr}}
\affiliation{Korea Astronomy and Space Science Institute 776, Daedeokdae-ro, Yuseong-gu, Daejeon, 34055, Korea}
\author{Yanghwan Kim\footnote{yanghwanphysics@snu.ac.kr}}
\affiliation{Korea Astronomy and Space Science Institute 776, Daedeokdae-ro, Yuseong-gu, Daejeon, 34055, Korea}
\affiliation{Department of Physics and Astronomy, Seoul National University 1 Gwanak-ro, Gwanak-gu, Seoul, 08826, Korea}
\begin{abstract}
	Although the Blandford-Payne process, the standard model for the production of AGN jet outflow, has been fully acknowledged and long-known in both the theoretical Astrophysics and observational Astronomy communities, subsequent research works to gain observational supports have been quite rare. In the present work, therefore, we would like to suggest a likely event and encourage its observation which demonstrates observational supports for the Blandford-Payne process. That is, we propose the coupling of it to the well-known Bardeen-Petterson effect. In order for this set-up to comply with our objective stated above, however, the two coupled processes need to be well-resolved. We, therefore, carefully study and present the condition for this to take place. We also point out that this major concern of our present work allows us to measure the strength of the intra-galactic magnetic field which has been known to be uneasy and unclear even in the galaxy observation Astronomy community for a long time.
\end{abstract}
\maketitle
\section{Introduction}
It is interesting to note that, although the Blandford-Payne process\citep{Blandford82} has been long known and acknowledged as the standard model for the generation of AGN jet outflow, subsequent research works to support its observational evidence has been quite rare. In the present work, therefore, we would like to suggest an event which demonstrates observational supports for the Blandford-Payne process. Indeed, it is quite hard to speculate any simple and direct observational support for the actual operation of the Blandford-Payne process. Therefore, in the present work, we would like to propose the coupling of the Blandford-Payne process to another event in such a manner that this supplementary event could demonstrate the actual operation of the Blandford-Payne process. Here, we employ the well-known Bardeen-Petterson effect\citep{Bardeen75} to play such a role. To be more concrete, in what follows, we demonstrate that by coupling these two events, the Bardeen-Petterson effect would shift or rotate the preferred direction in which the active galactic nuclei(AGN) jet outflow propagates. To this end, we should make sure that the events have to be well resolved in a time sequence. By requiring that the events have to be well resolved in a time sequence, we demonstrate the condition on the strength of the intra-galactic magnetic field. It is interesting to note that such direct measurement of the strength of the galactic magnetic field incidentally allows us to check the validity of the Blandford-Znajek mechanism\citep{Blandford77, Rees84, Macdonald82} for the central engine of AGN as it predicts that the AGN luminosity is given by $L_{BZ}\sim B^2M^2G^2/c^3$ which is to be compared with the observational data $L_{obs}\sim 10^{46}$erg s$^{-1}$.

\section{Generation of AGN Jet Outflow}
\subsection{The Blandford-Payne process}
Given the large-scale poloidal magnetic field, the Blandford-Payne process for the generation of AGN jet outflow could be described as follows.
\begin{itemize}
	\item[1.]On the disk, the outflow begins and reaches the poloidal component of magnetic field by gas pressure in a corona. 
	\item[2.]The flow is centrifugally driven out along the poloidal magnetic field line that makes an angle of less than $60\deg$ with respect to the disc surface.
	\item[3.]Far away from the disk, the toroidal component of magnetic field comes into play and collimates the outflow into a pair of anti-parallel jets moving perpendicular to the disk.
\end{itemize}
In this way, magnetic stress can extract the energy and angular momentum from an accretion disk independently of the presence of viscosity.
\subsection{Time scale for accretion-Jet flow}
As a simple estimate for $\Delta t_{BP}$, the time scale of the whole outflow process, would be given by;
\begin{equation}
	\Delta t_{BP}\sim\frac{l}{c_v},
\end{equation}
where the $c_v$ denotes the sound speed(Alfv\`ein velocity), and the $l$ denotes the length scale over which the poloidal magnetic field extends, which are given as
\begin{align}
	c_v &=\left(\frac{dp}{d\rho}\right)^{1/2}=\frac{B}{\sqrt{\mu_0\rho}},\quad\text{and}\\
	l&=10^3\frac{GM}{c^2},
\end{align}
respectively, for radio jet(or accretion disk) length scale of AGN, where $p$ is the gas pressure of plasma; $B$ is the magnetic flux; $\rho$ is the plasma density; $\mu_0$ is the vacuum permeability; $G$ is the Newton's gravitational constant; $M$ is the mass of the black hole, and $c$ is the speed of light.
\section{Precession of the accretion disk}
\subsection{The Bardeen-Petterson effect}
General relativistic magnetohydrodynamics treatment for the deformation of the accretion disk states that the Bardeen-Petterson precession of the accretion disk developes within the Bardeen-Petterson radius $r_{BP}$ which is given, rough, by;
\begin{equation}
	r_{BP}\gtrsim 10^2\frac{GM}{c^2},
\end{equation}
which reaches far outside the horizon.
\subsection{Time scale for accretion disk precession}
Since the total torque involved in the Bardeen-Petterson effect consists of three parts: alignment, precession, and spin-down components. Evidently, the precession time scale and the alignment time scale would be of the same order;
\begin{equation}
	\Delta t_{LT}\sim\frac{2\pi}{\Omega_{GM}},
\end{equation}
where
\begin{equation}
	\Omega_{GM}=\frac{2GJ}{c^2r^3_{BP}}\lesssim a_*\frac{2G^2M^2}{c^3\left(100GM/c^2\right)^3}=10^{-6}\frac{2c^3a_*}{GM};
\end{equation}
$J$ is the total angular momentum of the black hole which is given by $J=2GM^2a_*/c$, and $a_*$ is the specific angular momentum of the black hole.
\section{Concurrence of the Blandford-Payne process with the Bardeen-Petterson effect}
As we mentioned in the introduction, we would like to propose the coupling of the Blandford-Payne process to the Bardeen-Petterson effect. By coupling these two events, the Bardeen-Petterson effect would shift or rotate the preferred direction in which the AGN jet outflow propagates. We begin with a careful prerequisite condition for this to take place. That is, we should make sure that the events have to be well resolved in a time sequence. Along this line, we demand the following condition.

If $\Delta t_{BP}\lesssim\Delta t_{LT}$, the concurrence of Blandford-Payne process and Bardeen-Petterson effect could be resolved(FIG. 1). Given $B=1$mG and $M=10^5M_{\odot}$, the time scales are $\Delta t_{BP}\sim 14.5$ hours, $\Delta t_{LT}\sim 18$ days, which are reasonable time scale to observe two subsequent events: two Blandford-Payne processes before and after the Bardeen-Petterson effect. We therefore suggest the observation of radio spectrum with very high degree of resolution employing, say, the VLBI tool(FIG. 2). If such spectrum is actually observed, it would support our theoretical expectation: namely, the possible concurrence of the Blandford-Payne process with the Bardeen Petterson effect.

\begin{figure}[!ht]
	\centering
	\includegraphics[width=0.5\textwidth]{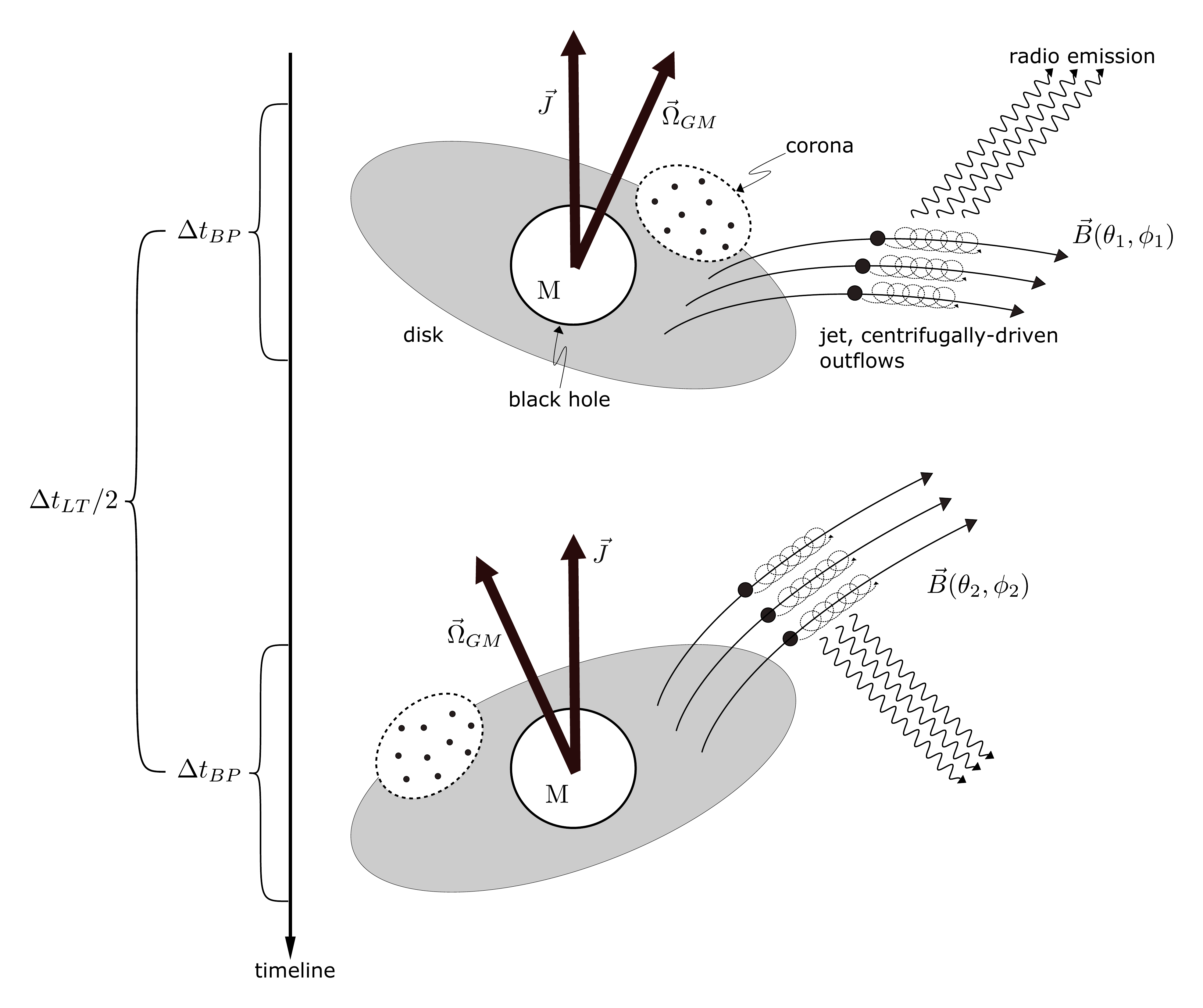}
	\caption{Cartoon for the concurrence of the coupled Blandford-Payne process and the Bardeen-Petterson effect. The $(\theta_1,\phi_1)$ and $(\theta_2,\phi_2)$ denote the initial and final angular position of the accretion disk. The $\vec{B}$ denotes the magnetic field at given angular position of accretion disk.}
	\includegraphics[width=0.5\textwidth]{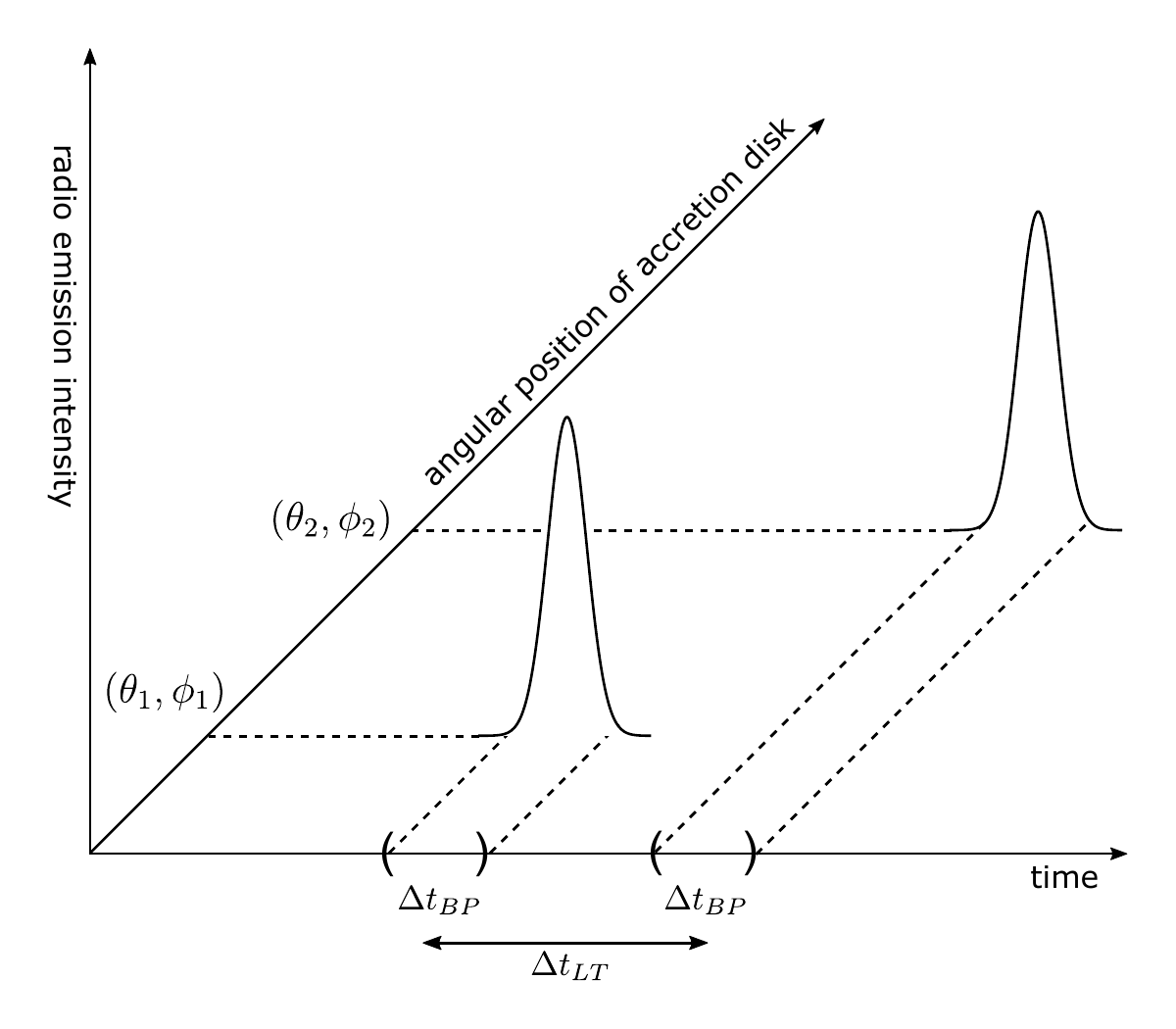}
	\caption{Schematic diagram for the Light Curve of the coupled Blandford-Payne process and the Bardeen-Petterson effect.}
\end{figure}

If such resolution of the two subsequent radio emission could be observed in the VLBI observation of an AGN galaxy sample, it implies that the strength of the galactic magnetic field of the corresponding AGN galaxy would be estimated to be $34\mu\text{G}\lesssim B$. It is interesting to note that such direct measurement of the strength of the galactic magnetic field incidentally allows us to check the validity of the Blandford-Znajek mechanism(Blandford \& Znajek 1977, Macdonald \& Thorne 1982, Martin 1984) for the central engine of AGN as it predicts that the AGN luminosity is given by $L_{BZ}\sim B^2M^2G^2/c^3$, which is to be compared with the observational data $L_{obs}\sim10^{46}$erg s$^{-1}$.

\end{document}